\documentclass[iop]{emulateapj}
\shorttitle{The Unimodal Distribution Of The BSS in M75} 
\shortauthors{Contreras Ramos et al.}

\begin{document}

\title{The Unimodal Distribution Of Blue Straggler Stars in M75 (NGC~6864)}

\author{R. Contreras Ramos, F. R. Ferraro, E. Dalessandro, B. Lanzoni}
\affil{Dipartimento di Astronomia, Universita di  Bologna, via Ranzani 1, 40127,Bologna, Italy}
\email{rodrigo.contreras@oabo.inaf.it}
\and
\author{R. T. Rood}
\affil{Astronomy Department, University of Virginia, P.O. Box 400325, Charlottesville, 
VA 22904, USA}

\begin{abstract}
We have used a combination of multiband high-resolution 
and wide-field ground-based observations to image the 
Galactic globular cluster M75 (NGC~6864).  
The extensive photometric sample covers the entire cluster extension, from 
the very central regions out to the tidal radius, allowing us to determine 
the center of gravity and to construct the most extended star density 
profile ever published for this cluster. We also present the first 
detailed star counts in the very inner regions.  
The star density profile is well re-produced by a standard King model 
with core radius $r_c\sim 5.4^{\prime\prime}$ and intermediate-high 
concentration $c\sim 1.75$. The present paper presents a detailed study 
of the BSS population and its radial distribution. 
A total number of 62 bright BSSs (with $m_{\rm F255W}\lesssim 21$, 
corresponding to $m_{\rm F555W} \lesssim 20$) has been identified, and 
they have been found to be highly segregated in the cluster core. 
No significant upturn in the BSS frequency has been observed in the 
outskirts of M75, in contrast to several other clusters studied with the 
same technique. This observational fact is quite similar to what has 
been found in 
M79 (NGC~1904) by \cite{lanzoni07a}. Indeed the BSS radial distributions
in the two clusters is qualitatively very similar, even if in M75 the relative 
BSS frequency seems to decrease significantly faster than in 
M79: indeed it decreases by a factor of 5 (from 3.4 to 0.7) 
within 1 $r_c$. Such evidence indicate that the vast majority
of the cluster heavy stars (binaries) have already sunk to the core.  

\end{abstract}


\section{Introduction}
\cite{sandage53} first identified a bizarre group of
apparently massive young stars immersed in the $\sim13$ Gyr old 
population of the globular cluster (GC) M3. 
They appeared as an extension of the classical hydrogen-burning main 
sequence (MS), both hotter and more luminous than the
stars that define the MS turnoff (TO) point. This was the reason 
why they were properly named blue straggler stars (BSSs). 
Being devoid of gas after the burst that formed the bulk of
stars, GCs would be unable to form such ``young'' objects, and their existence 
in GCs contradicts the current picture of stellar evolution, since massive 
stars should have evolved into white dwarfs long ago. 
Their nature has been puzzling for many years, and even today their 
formation mechanism is not completely understood.
Nowadays, BSSs are considered objects more massive than the normal
MS stars \citep[$M \sim 1.2M_{\odot}$, see][]{shara97,ferraro06a} 
that somehow have increased their initial mass.
Two explanations for the generation of BSSs are currently leading:
mass exchange in primordial binary systems \citep[][]
{mccrea64,zinn76,knigge09}, and stellar mergers induced
by collisions in dense environments \citep[][]{hills76,leonard89}. 
The two scenarios do not necessarily
exclude each other \citep[][]{bailyn92} and might co-exist 
in the same GC \citep[see the case of M30,][]{ferraro09}.

Gravitational interactions between stars leads to the 
dynamical evolution of GCs on timescales generally smaller than their
ages \citep[][]{meylan97}. 
One signature of such evolution is mass segregation: over time, 
most massive stars slow down and sink to the cluster core, while lighter 
stars acquire speed and move to its periphery. 
Thus, BSSs are expected to preferentially populate the innermost region of star
clusters.
Because of the stellar crowding, the acquisition of complete samples of BSSs
in the core of GCs is a quite difficult task in the optical bands. 
Conversely, it is easy in the UV bands \citep[][]{paresce91}. 
In fact,
due to their high temperatures they are especially bright in the UV, while 
cool populations such as red giant branch (RGB) stars appear to be quite 
faint \citep[][and references therein]{ferraro97}.
In addition, thanks to the advent of the {\it Hubble Space Telescope} (HST) 
coupled with wide-field imagers on ground--based telescopes,
it has become possible to survey the BSS population over the entire
extension of GCs.
Taking advantage of these possibilities, the BSS radial distribution 
in several GCs has been already carefully analyzed 
\citep{ferraro97,ferraro99,ferraro03a}. 
The common feature 
in most of them is that BSSs show a bimodal radial distribution: they 
appear to be strongly concentrated in the central regions, at 
intermediate radii the distribution decreases down to a minimum value and 
then it increases again at outer radii. 
This behavior has been interpreted in terms of mass segregation by
\cite{mapelli04,mapelli06}. Notable exceptions 
to this rule are $\omega$ Centauri 
\citep{ferraro06a}, NGC~2419 \citep{dalessandro08} and, 
the recently studied, 
Pal~14 \citep{beccari11}, which show a ``dynamically young'' 
state of evolution with the BSS population being 
homogeneously distributed across 
the entire cluster extension, showing no evidence of central segregation.
A different behavior has been found in the GC M79 \citep{lanzoni07a},
which instead shows a large concentration of BSSs within the core 
but no evidence of an upturn in the outerparts of the cluster. 
This would suggest that the action of mass segregation on the current BSS 
population is already effective at any distance from the cluster center.

Here we present a complete investigation of the BSS population
M75 (NGC 6864). This is a massive and distant GC 
\citep[$M_{\rm V}=-8.57$, $d \backsimeq 20.9$ kpc;]
[2010 Edition, hereafter HA96]{harris96} 
with a moderately high metallicity [Fe/H] $\sim 1.29$
\citep{carretta09}, 
appearing to have a trimodal HB \citep{catelan02}
and with a relatively high central density, 
log$\rho_{0}/(M_{\odot}$ pc$^{-3})=4.9$ \citep{pryor93}
in which BSSs have not been already properly studied. 
The only previous work, in which BSSs were briefly analyzed was 
realized by \cite{catelan02} as part of their 
analysis of the overall properties of the 
Color Magnitude Diagram (CMD) of M75 using ground-based data.  
The plan of the paper is as follows. In section~2 the photometric data set
is presented and discussed. In section~3 we present the determination of
the center of gravity and the star density profile of the cluster. 
In section~4 we discuss the CMD selection and the properties of the BSS sample.
In section~5 we draw our conclusions.

\section{Data--sets and Analysis}
\begin{figure*} [!t]
\centering
\includegraphics[angle=180,width=17cm]{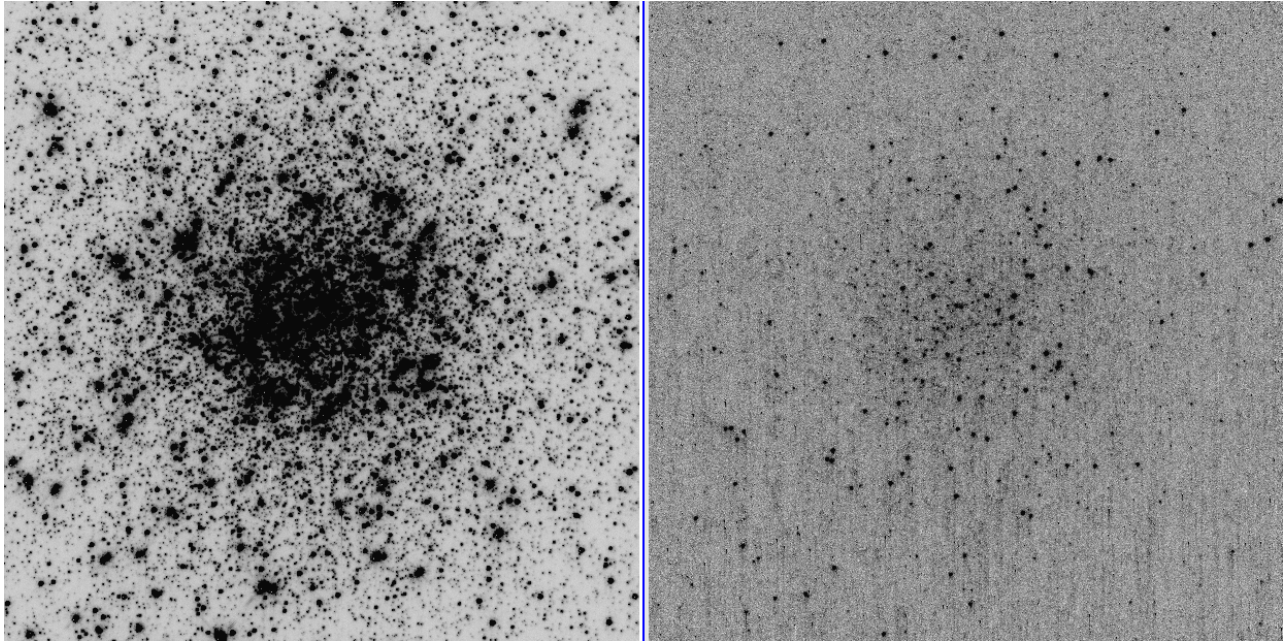}
\caption[]{The cluster center as imaged by the PC ($35^{\prime\prime}
\times 35^{\prime\prime}$) of the HST/WFPC2 through the F555W (left panel) 
and the (UV) F255W (right panel) filter. The optical image is dominated by 
cool bright red giants which blend together preventing the measure of 
hot (faint) objects. In the UV image, the brightest objects are HB stars 
and BSSs and blending effects
are not present even in the very central region of the cluster. In this image,
hot objects as HB and BSSs are easily identifiable and accurately measurable.}
\label{images}
\end{figure*}
\subsection{Observations}
The present analysis is based on a combination of two sets of
observations:

{\it The High Resolution (HR) data set.---} The inner and most crowded 
region of the cluster has been sampled using the exquisite quality 
images from the Wide Field Planetary Camera 2 (WFPC2) on board the HST.
These data were obtained under program GO 11975 
(P.I. F. Ferraro) on March 2009 as part of a major project aimed 
to census the UV sources in GCs.
The cluster was roughly centered on the planetary camera (PC) of the 
WFPC2 that provides better spatial resolution 
($\sim 0^{\prime\prime}$.046 pixel$^{-1}$), while
the lower resolution  ($\sim 0^{\prime\prime}$.1 pixel$^{-1}$) 
wide--field cameras sample the surrounding regions of the cluster. 
The WFPC2 has a total field of view (FOV) of 
150$^{\prime\prime} \times 150^{\prime\prime}$, which 
corresponds to a projected area of $15\times15$ pc$^{2}$ at the distance
of M75. 
The observational material consists of a series of images in 3 carefully 
selected filters: F255W (4$\times$1200 s), F336W (3$\times$700+1$\times$100 s)
and F555W (3$\times$100+1$\times$5 s). 
Both the high angular resolution and the UV sensitivity of HST are essential
to identify the BSSs among the much more luminous 
(in the classical visible bands) red giants belonging to 
the cluster. Fig.~\ref{images} shows the clear advantage of using UV images 
to search for hot--objects: it compares two images of the cluster center 
taken through the F555W filter (left panel) and through the F255W 
filter (right panel). 
As can be appreciated, the optical image is dominated by the emission 
of bright, red (cool) giants, that blend 
together due to the high crowding conditions, making difficult
(if not impossible) the secure identification of BSSs. Conversely, in the UV 
image (right panel) RGB stars are much dimmer and hot objects like
horizontal branch (HB) stars and BSSs, are the brightest sources and 
therefore easily distinguishable even in the cluster center.

\begin{figure}
\begin{center}
\includegraphics[width=8.5cm]{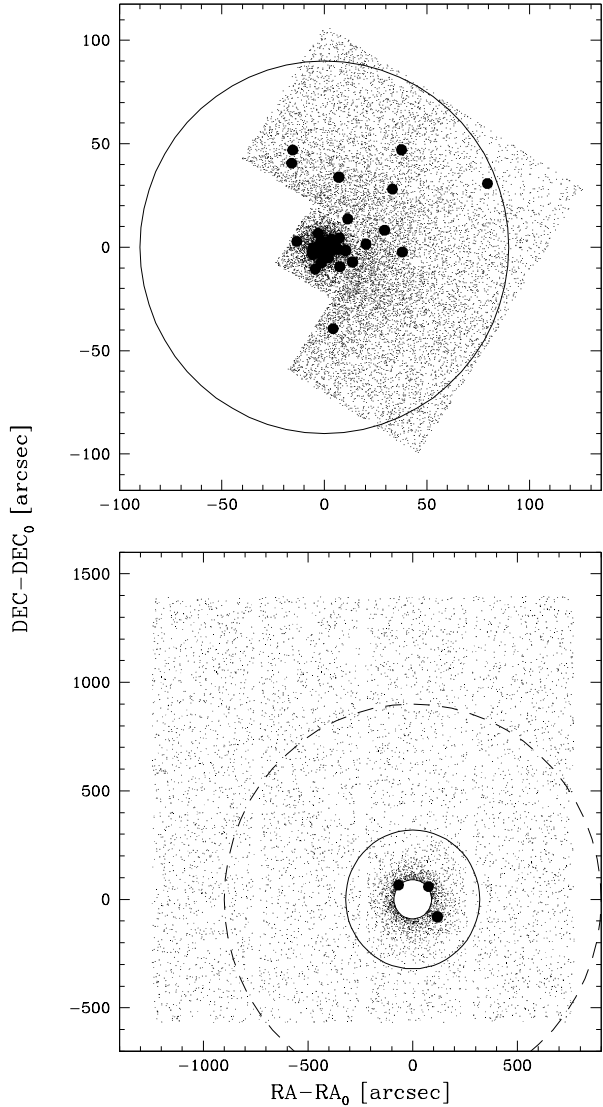}
\caption[]{Upper panel: Map of the HR sample. The positions of 
all the measured stars are plotted
with respect to the center of gravity. The BSSs identified
in this sample are marked as large filled circles. The circle at 
$r=90^{\prime\prime}$ from the cluster center marks the edge of the sample
used in the present snalysis.
Lower panel: Map of the WF sample. Large filled circles are the identified 
BSSs. The solid circle
at $r=5^{\prime}$ marks the position of the tidal radius. The dashed circle
at $900^{\prime\prime}$ from the cluster center delimitate the region 
adopted to evaluate the background contamination.} 
\label{fov}
\end{center}
\end{figure}

{\it The Wide Field (WF) data set.---}The HST/WFPC2 data were 
complemented with a set of WF images acquired in the $V$ and $I$ 
bands during an observing run at the 2.2 m ESO-MPI telescope at 
ESO (La Silla) in 2002 June using the Wide Field Imager (WFI). 
With a global FOV of $34^{\prime} \times 33^{\prime}$ provided by 
a mosaic of eight CCD chips (each chip compose of 2050 $\times$ 4100 pixels 
with a pixel scale of about $\sim 0^{\prime\prime}$.238 pixel$^{-1}$) this 
data covers by far the entire cluster extension (see bottom
panel og Fig.~\ref{fov}), 
allowing exceptional imaging of the outer and less crowded portion of the 
cluster, which was roughly centered on chip 7. 

\subsection{Photometry} 
Each WFPC2 frame was processed through the standard HST-WFPC2 pipeline
for bias subtraction, dark correction and flat-fielding. Then, the
single--chips were extracted from the 4--chip mosaic, and
analyzed separately. Due to vignetting problems affecting the borders of the
single--chips, the pixel regions x $\leq 50$; y $\leq 50$ have been excluded 
from the analysis.
Frames obtained with gain = 7e/adu (F225W) and 15 e/adu (F336W, F555W) were
available, so care was taken to adjust the various parameters to the 
applicable gain value for each frame. 
The brightest, non saturated stars were selected
among those having no nearby companions or defects within a few pixels,
to construct an analytical Point Spread Function (PSF). 
In order to fit the star brighteness profile in each frame we adopted
a two-component PSF, obtained by combining: (a) an analytical component 
reproduced by a \cite{moffat69} function whose shape has been allowed to vary
linearly with the star position in the frame and (a) a numerical component 
which allows to take into account for the systematic difference between the 
observed star profile and the analytic approximation.
In long exposures, especially in  
the UV band, a huge amount of cosmic rays were present, making the PSF 
construction tricky. Accordingly, we combined the UV--images using the 
IRAF task imcombine (imposing $\it{crreject}$ to eliminate cosmic rays) 
in order to selected the PSF star candidates in a decontaminated median frame.
The stellar photometric reduction was carried out using the
DAOPHOTII/ALLFRAME package \citep[][1994]{stetson87}. 
A preliminary photometry was performed
in order to construct a list of stars and obtain accurate coordinates for 
each single frame. We used DAOMATCH/DAOMASTER to determine coordinate
transformations for every image to a carefully selected reference one. These
transformations were then used by the MONTAGE task routine 
to create a ``master'' frame combining the F255W and F336W data in
order to eliminate cosmics rays, hot pixels and other spurious sources, and 
obtain a high signal/noise image for star finding. 
In our opinion, this filter selection is the best 
compromise for the detection of both hot and cold sources. 
We used the DAOPHOT/FIND routine to selected objects in the master image 
imposing a detection limit $\geq 3 \sigma$ and then the entire star list
was given as input to ALLFRAME, which performed a simultaneous 
PSF--photometry of all the individual frames using the coordinate 
transformation among all the images. 
Using again DAOMASTER, first, we created a
catalog of mean magnitudes of each filter, and then we combined these 
individual catalogs to obtain a final master one in which the star list 
includes all the sources detected in at least two filters. As a final step, we 
transformed the F255W, F336W and F555W instrumental magnitudes into the 
VEGAMAG photometric system using the procedure described in 
\cite{holtzman95} with the zero points listed in Table 5.1 of the HST
Data Handbook.

As far as the reduction of the WFI data is concerned, the images were
initially corrected for bias and flat-field with the standard IRAF routines. 
Then stellar photometry was obtained running DAOPHOTII/ALLFRAME on all the 
images simultaneously, as describe above.
The calibration of the instrumental magnitudes into the VEGAMAG 
photometric system has been done in 2 successive steps: First, since no 
photometric standard stars were available, we used the calibrated
catalog obtained by \cite{catelan02} to link the WFI instrumental 
magnitudes to the standard Johnson 
system. With this purpose in mind, we performed a cross correlation
between the two catalogs, and selected about 700
stars in common. This sample covers a sufficiently wide range in color to
prevent any residual, uncorrected color trend. These selected stars
were then used to calibrate the WFI data by means of a least-squares fit.  
Second, we applied \cite{holtzman95} equations 7 and 9
(taking as zero points those presented in their Table 9) to convert  
the magnitudes from the Johnson to the VEGAMAG system, thus making
the WFI dataset photometrically homogeneous with the HST one. 
In the following, we will therefore adopt the notation $m_{\rm F555W}$ and 
$m_{\rm F814W}$ to label also the V and I magnitudes of the WFI dataset.

\subsection{Astrometry}
In order to place the HR and WF samples in a common 
coordinate system we have used astrometric standard stars selected from 
the new Guide Star Catalog (GSCII). 
The procedure has been described in previous papers \citep{ferraro03b}. 
We summarize here the main steps: 
using several hundreds of stars in common between the WFI catalogue
and the GSCII, the stars position in each of the eight WFI chips were placed 
on the absolute astrometric system. For this purpose, we have used CataXcorr, 
a program developed at the Bologna Observatory (P. Montegriffo, private
comunication), 
to perform roto-translation procedures and allow accurate absolute
positioning of the stars. At the end of the procedure, the position
residuals were of the order of $\sim 0^{\prime\prime}$.2 in both 
right ascension ($\alpha$) and declination ($\delta$). 
Then, the HST catalog was placed on the absolute astrometric system 
by cross-correlating a few hundred stars in common 
between the WFI and the WFPC2 FOVs (see Fig.~\ref{fov}).
We estimate that the astrometric uncertainty for the WFPC2 stars is less than
$\sim 0^{\prime\prime}$.2 in both $\alpha$ and $\delta$.  

\section{Results}
\subsection{Center of Gravity}
Taking advantage of the knowledge of the exact positions of the stars,
even in the innermost central regions, we have estimated the geometrical
center ($C_{\rm grav}$) of the star distribution as the barycenter 
of the within the PC FOV. In doing so, 
we have performed the iterative procedure described in \cite{montegriffo95}. 
In order avoid incompleteness effects and possible
statistical fluctuations, we considered stars contained within 
four circular areas with
different radii ($10^{\prime\prime}$, $11^{\prime\prime}$, $12^{\prime\prime}$
and $13^{\prime\prime}$) and three different magnitude cuts
($m_{\rm F555W}=20.5, 21$ and $21.5$), adopting as first 
guess the center reported by HA96. Finally, we simply took the 
average of the twelve computed values as the best estimate of $C_{\rm grav}$ 
position, which is located at 
$\alpha_{\rm j2000}=$20$^h$ 06$^m$ 4$^s$.85, 
$\delta_{\rm j2000}=$-21$^{\circ}$ 55$^{\prime}$ 17$^{\prime \prime}$.85. 
This new value for the barycenter of M75 differs slightly  
($\Delta\alpha=0.16^{s}$, $\Delta\delta=-1.65^{\prime \prime}$) 
from the previous one listed in HA96 based on the surface brightness
distribution.

Finally, the two catalogs were merged together. Being aware of the superior
resolution capabilities of HST and the high effectiveness of the UV 
observations in detecting BSSs in the crowded central region of the cluster,
we have considered only stars measured in the HR sample at 
$r \leq 90^{\prime\prime}$ from $C_{\rm grav}$, and we have consistently
restricted the WF sample to the region $r > 90^{\prime\prime}$.
Because of the peculiar shape of the WFPC2 FOV, our choice implies 
that part of the area in the inner $90^{\prime\prime}$ (see top panel of 
Fig.~\ref{fov}) is covered by neither of the two samples. 
However, at the same time, this selection minimizes incompleteness effects 
and stellar blends in the central region of the cluster. 
\begin{figure}
\begin{center}
\includegraphics[width=9.0cm]{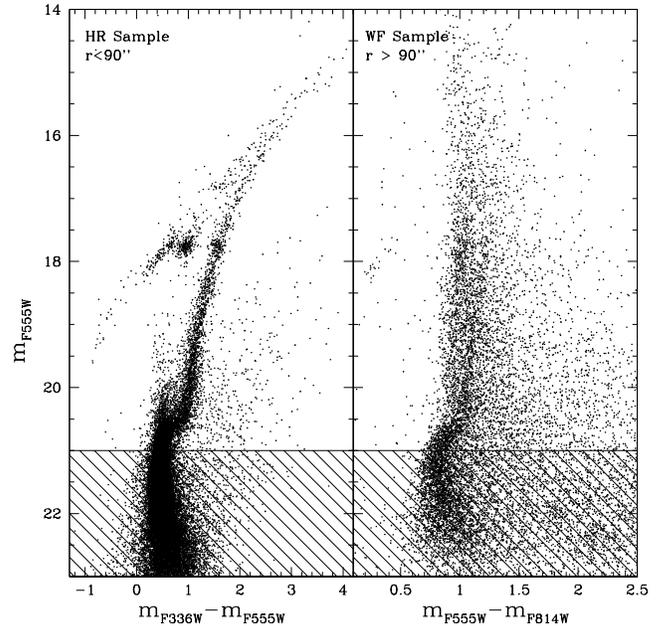}
\caption[]{The optical CMDs for the HR and WF samples. The shaded region
($m_{\rm F555W} >$ 21) delimit the samples excluded for the 
computation of the star density profile.}
\label{cmdhstwfi}
\end{center}
\end{figure}
At the end of the whole procedure, we ended up with a final catalog fully 
homogeneous both in magnitudes and coordinates, covering the 
entire cluster extension. Fig.~3 shows the optical  
($m_{\rm F555W}$, $m_{\rm F336W}-m_{\rm F555W}$) and ($m_{\rm F555W}$, 
$m_{\rm F555W}-m_{\rm F814W}$) CMDs for the HR and WF samples, respectively. 
A total of 44220 stars are plotted:
30806 in the HR sample and 13414 in the WF sample, respectively.
Because of its large extension
(reaching $r > 25^{\prime}$ from the cluster center) the WF sample is 
dominated by the Galactic field. These samples have been first used
to compute the star density profile.  

\subsection{Density profile} 
The dynamical status of a GC can be probed using surface
brightness and/or density stellar profiles. However, when evaluating the 
brightness over small regions, the  profiles
may be affected by large fluctuations caused by few bright stars. On the
other hand, we can take advantage of the fact that we resolve stars even
in the innermost region of the cluster and directly build the star
density profile, which is likely the most robust tool for determining
the cluster structural parameters \citep{lugger95}. 
\begin{figure}
\begin{center}
\includegraphics[width=9.5cm]{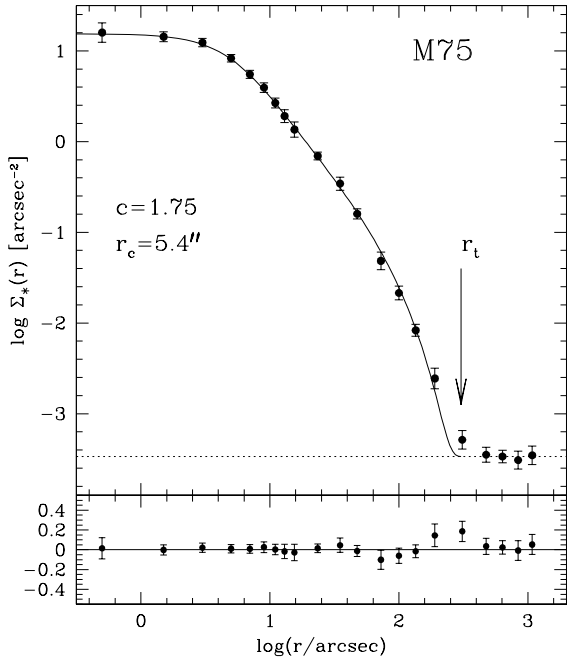}
\caption[]{Observed surface density profile (dots) and best--fit King model
(solid line). The dotted line indicates the measured level of the background.
The parameters of the King model (core radius and concentration) are indicated 
in the Figure. The location of the cluster core radius 
(at $r_{\rm t}=5^{\prime\prime}$)
is marked by the vertical arrow. The lower panel shows the residuals of the
fitting procedure in each radial bin.}
\label{densprof}
\end{center}
\end{figure}
Using direct counts, we have therefore 
determined the projected density profile of M75 for its entire radial 
extension, from $C_{\rm grav}$ out 
to $\sim 1000^{\prime\prime}$ $(\sim100$ pc). 
To avoid incompleteness effects and keep field contamination 
as low as possible, we have 
considered only stars brighter than  $m_{\rm F555W}=21$ and located
within a $3\sigma$--width selection boxes running along the main
evolutionary sequences defined in the CMD by the cluster stars.
Following the procedure described by \cite{ferraro03b}, we have
divided our catalog into 22 concentric annuli centered on $C_{\rm grav}$
and each annulus, in turn, has been split in a number of equivalent
subsectors. The
number of stars lying within each subsector has been counted, averaged and
finally divided by the area of the subsector to eventually obtain the
mean star density of each annulus. The standard deviation between 
the subsectors in each annulus has been used as the uncertainty of the
star density. 
The radial density profile thus derived is plotted in 
Fig.~\ref{densprof} (solid dots), with the abscissas corresponding 
to the mid-point of each radial bin. The plot clearly shows that the star 
counts flatten beyond $r \gtrsim 400^{\prime\prime}$. This feature has
been used to directly estimate the contribution of the background level, which
is shown by the dashed line in Fig.~\ref{densprof}.
The adopted lower magnitude limit for the density profile counts 
($\sim$ MS TO level) imply that all the stars considered in the analysis 
have approximately the same mass. Accordingly, a mono-mass, isotropic King
model has been computed in order to reproduce the observed profile
(solid line in Fig.~\ref{densprof}). The best fit to our data  
provides a core radius $r_{c}=5.4^{\prime\prime}$ (which corresponds 
to $\sim0.5$ pc) and a concentration parameter $c=1.75$, which would
suggest that M75 has not already experienced the core collapse 
\citep[$c \gtrsim 2$,][]{meylan97}. 
Our results are in perfect agreement with those quoted by 
\cite{mclaughlin05}, $r_{\rm c}=5.4^{\prime\prime}$ and  $c=1.8$, both derived 
from the surface brightness profile.

\begin{deluxetable*} {lllllllc} [!t]
\footnote{The complete version of this table is available in the on line 
journal.} 
\tablecaption{The BSS Population of M75}
\tablewidth{0.0pt}
\tablehead{$Name$ & $R.A.$ & $Dec.$ & $Dist.$ & $F255W$ & $F336W$ & $F555W$ & $F814W$\\ &(degree)&(degree)&(arcsec)\\}
\startdata
BSS1  &301.5202501 & -21.9217358 &   0.401 &21.150 & 20.279 & 20.002 & ..... \\
BSS2  &301.5201192 & -21.9215229 &   0.520 &20.418 & 19.653 & 19.198 & ..... \\
BSS3  &301.5200227 & -21.9216803 &   0.712 &20.343 & 19.627 & 19.276 & ..... \\
BSS4  &301.5200460 & -21.9217440 &   0.740 &21.259 & 20.092 & 19.601 & ..... \\
BSS5  &301.5202849 & -21.9213937 &   0.859 &21.304 & 20.198 & 19.573 & ..... \\
BSS6  &301.5198820 & -21.9213940 &   1.426 &20.827 & 20.046 & 19.730 & ..... \\
....
\enddata
\label{bsstable}
\end{deluxetable*}

\subsection{Cluster population selection}
The first step to study the projected BSS radial distribution is the
appropriate selection of the BSS and the reference populations. 
In defining the population samples, here we followed the approach
already adopted in previous papers of this series 
\citep[e.g.][]{ferraro03a,lanzoni07a,dalessandro09}. 
We underline that the exact shape of the population selection  
boxes could slightly vary from cluster to cluster, depending on many
factors such as 
(a) the  cluster properties (for instance, the HB morphology and
presence of RR Lyrae, which, when observed at random phases, could be 
locate above or below the HB level an possibly contaminate 
the brightest portion of the BSS  population), 
(b) the quality of the photometry (which can significantly spread the 
sequences at the faint end), 
(c) the necessity to exclude field stars.
These differences, however, do not affect the comparison of the population 
radial distribution.
   
\subsubsection{BSS population}
\begin{figure}
\begin{center}
\includegraphics[width=8.5cm]{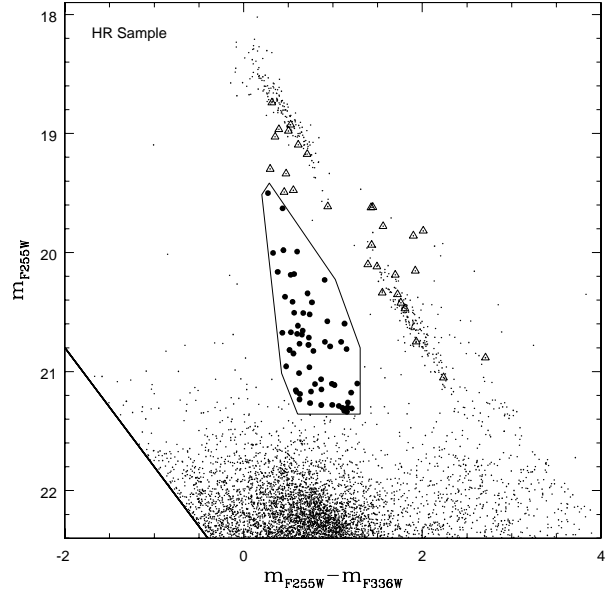}
\caption[]{Ultraviolet ($m_{\rm F255W}$, $m_{\rm F255W}-m_{\rm F336W}$) CMD 
for the HR sample. 
The solid line delimitate the detection limit at $m_{\rm F336W} \sim 22.8$.
The selection box used for the definition of the BSSs (plotted as large
filled circles) is shown. All the RR Lyrae stars identified in the HR field
are marked with large triangles.}
\label{cmduv}
\end{center}
\end{figure}

\begin{figure}
\begin{center}
\includegraphics[width=8.5cm]{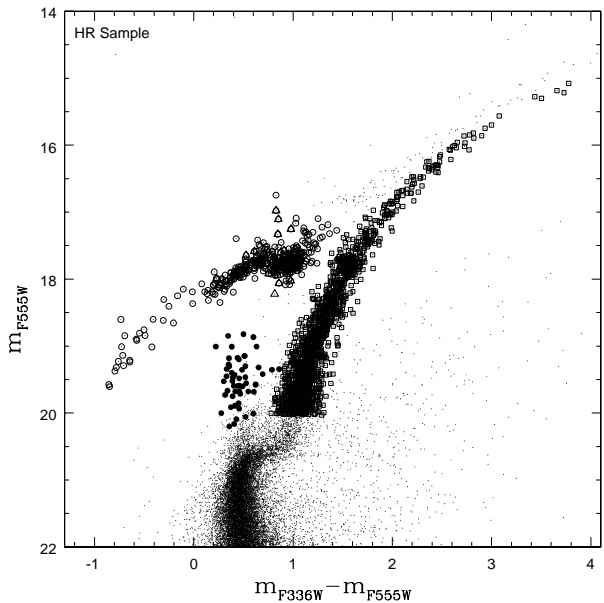}
\caption[]{Optical ($m_{\rm F555W}$, $m_{\rm F336W}-m_{\rm F555W}$) CMD 
for the HR sample.
The selected populations are plotted with different symbols: RGB (open 
squares), HB (open circles), known RR Lyrae (open triangles). The BSSs selected
from the UV CMD shown in Fig.\ref{cmduv} are plotted as filled circles. Note
the presence in the BSS region of spurious objects probably due to optical 
blends of TO stars.}
\label{cmdopthst}
\end{center}
\end{figure}

Previous works aimed at studying the BSS population in GCs
\citep[see][and references therein]{ferraro03a,lanzoni07a,dalessandro09} have
shown that BSSs are easily distinguished from the cooler stars of the
turnoff, SGB and RGB when using UV CMDs. The main reason is that at these 
wavelengths the cluster light is dominated by hot stars, especially the 
blue HB and the BSSs. 
As shown in Fig.~\ref{images}, HB stars and BSSs are the 
brightest objects appearing in the UV image, while the contribution from
cooler MS, RGB and SGB stars is almost negligible. 
This is even more clear in the UV CMD shown in Fig.~\ref{cmduv}. 
Indeed, the location and morphology of the main evolutionary branches in 
the UV bands are very different from the classical optical CMD 
(see Fig.~\ref{cmdopthst}). 
The RGB is very faint and the HB appears as a narrow branch crossing
diagonally the CMD.
In turn, the BSS population defines a nearly vertical sequence spanning
$\sim 3$ mag in $m_{\rm F255W}$. They are clearly separable from the fainter 
TO and SBG stars. For these reasons our primary criterion for the BSS
selection is based on the position of stars in the 
($m_{\rm F255W},m_{\rm F255W}-m_{\rm F336W}$) plane. 
However, an unequivocal definition of the faint edge of the BSS population is 
not simple since the BSS sequence merges smoothly into the MS--TO region 
without showing any gap or discontinuity. 
To avoid any possible contamination due to
blends, incompleteness and TO and SGB stars in 
the BSS sample, here we limit our
selection criteria to the brightest portion of the BSS sequence. Thus, we 
have imposed $m_{\rm F225W} \leq 21.3$ as the fainter threshold of the BSS 
selection box. This limit is nearly one mag brighter than the MS TO in the
UV plane. 
The adopted selection box for the BSSs in the HST sample is shown in 
Fig.~\ref{cmduv}. 
A total of 59 bright-BSSs have been counted in this way.
In Fig.~\ref{cmdopthst} the BSSs selected in the UV plane are plotted
in the optical CMD (filled circles). This, once again, illustrates the 
problem of using optical CMDs to identify BSSs in crowded regions.
Even with the high resolution achieved with HST, the BSS region
can be populated by spurious objects as a result of blending in the 
dense GC cores. 
In fact, there are many objects lying in the BSS region
that were not classified as genuine BSSs in the UV diagram. We believe
that most of them may be considered stellar blends. In the
same vein, there are some stars lying very close to the RGB sequence,
illustrating that the optical magnitudes may suffer blend/crowding
problems. 

In order to define a selection box for BSSs in the WF sample,
we have adopted the same $m_{\rm F555W}$ magnitude range 
($18.5 \lesssim m_{\rm F555W} \lesssim 20.2$)  
as for the HST (Fig.~\ref{cmdopthst}), while the 
red edge has been conservatively chosen in order to limit the effect
of the field contamination.
Three BSSs have been selected in this way (see Fig.~\ref{cmdoptwfi}).   
A detailed comparison  with the sample of 26 BSS found by \cite{catelan02}
in M75 was performed. Six of their candidates are outside the region sampled 
by our observations: they lie in the innermost region ($r<90"$) not covered by 
the WFPC2 FoV (see top panel of Fig.~\ref{fov}). 
Of the eight stars found in the area covered 
by our HST data, only one lies in the BSS selection box. The other seven 
are SGB, RGB stars well below the BSS sequence in the 
($m_{F255W}, m_{F255W}-m_{F336W}$) plane. Because of the higher HST spatial 
resolution and the BSS selection performed in the UV plane, we suggest that 
these stars are indeed blends in the optical bands. The 
remaining 12 candidates are located in the external  region (with $r>90''$): 
nine of them appears to be SGB or TO stars, well below the selection 
threshold.  One candidate has a magnitude consistent with our magnitude 
range  but it has a color too red to be considered a genuine BSS. Three 
candidates may be faint-BSS, since they lie along the BSS sequence but they 
are fainter than our threshold magnitude.  
In summary, a total of 62 BSSs has been selected in our field 
of M75, and all of them have been confirmed by visual inspection.
The identification, coordinates, distance from the derived $C_{\rm grav}$ 
and magnitudes of all the selected BSSs are 
listed in Table~\ref{bsstable}, which is available in full size
in the electronic version of the journal.
The spatial distribution of the selected BSSs is shown in Fig.~\ref{fov}. 
The BSS distribution appears
to be highly concentrated within the innermost region of the cluster. 
In particular, most of the BSS candidates lie within the 
inner $\sim 15^{\prime\prime}$ (PC FOV).

\subsubsection {Reference Populations}
\begin{figure}[!b]
\begin{center}
\includegraphics[width=8.5cm]{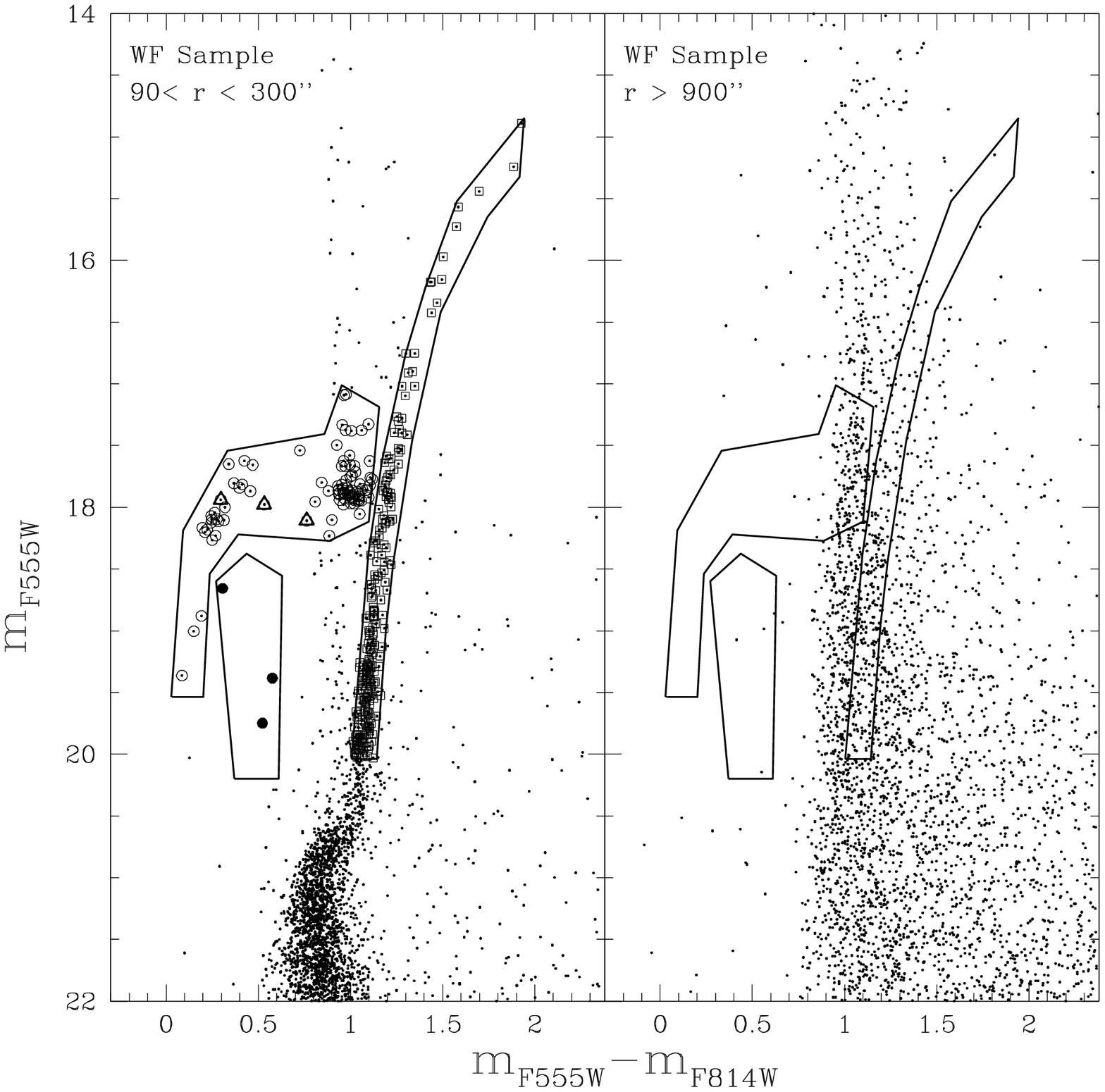}[!b]
\caption[]{Left panel: CMD of the inner $300^{\prime\prime}$ of the WF 
sample. The selection boxes used to define the BSS and RGB samples are 
plotted. Right panel: CMD of the field region, corresponding to the most 
external region ($r > 900^{\prime\prime}$) of the WF sample (see 
Fig.~\ref{fov} lower panel), with the population selection boxes superimposed.}
\label{cmdoptwfi}
\end{center}
\end{figure}
In order to quantify this visual impression
we need to compare the BSS distribution to that of a 
reference population, which is expected to display a non--peculiar radial
trend within the cluster. As in previous
papers \cite[see][]{ferraro03a, dalessandro08}, we have
adopted the RGB and HB stars as reference populations.
The selection of the RGB stars has been performed in the optical
($m_{\rm F555W},m_{\rm F336W}-m_{\rm F555W}$) 
and ($m_{\rm F555W},m_{\rm F555W}-m_{\rm F814W}$)
planes, for the HR and WF samples, respectively. For both samples
a magnitude threshold at $m_{\rm F555W} \lesssim 20$ has been adopted. The 
selected RGB stars are shown as open squares in Fig.~\ref{cmdopthst} and
\ref{cmdoptwfi}. 
The color limits of the boxes have been chosen in order to follow the
morphology of the RGB in each plane and in order to minimize the contamination
from AGB stars and field outliers. Slightly different 
assumptions on the selection boxes would have included or excluded a few 
objects without affecting the results of this analysis.  
Following this approach, we selected 1434 RGB in the HR sample and 241 RGB 
in the WF sample within $r < 300^{\prime\prime}$, this radial limit 
corresponding to the estimated tidal radius of M75 (see Sect. 3.2). 
The HB stars have been selected
essentially in the UV plane for the HR sample, and a careful analysis has been
performed in order to check that all these objects
lay also on the HB in the optical plane. The final selection is
shown in Fig.~\ref{cmdopthst} where HB stars are plotted as large open
circles. The same symbols are used to mark the HB population 
in the WF sample (left panel of Fig.~\ref{cmdoptwfi}), which has been 
selected using the optical ($m_{\rm F555W},m_{\rm F555W}-m_{\rm F814W}$) plane.
We carefully identified all the known RR Lyrae stars 
\citep{corwin03} lying in our FOV. 32 RR Lyrae stars have been found 
and they are marked as open triangles in Fig.~\ref{cmduv}, \ref{cmdopthst}, 
and \ref{cmdoptwfi}.
The total number of HB stars is 425 (29 of them are RR Lyrae) for the HR 
sample and 95 (3 RR Lyrae) for the WF sample ($r < 300^{\prime\prime}$). 

\begin{figure} [!b]
\includegraphics[width=8.5cm]{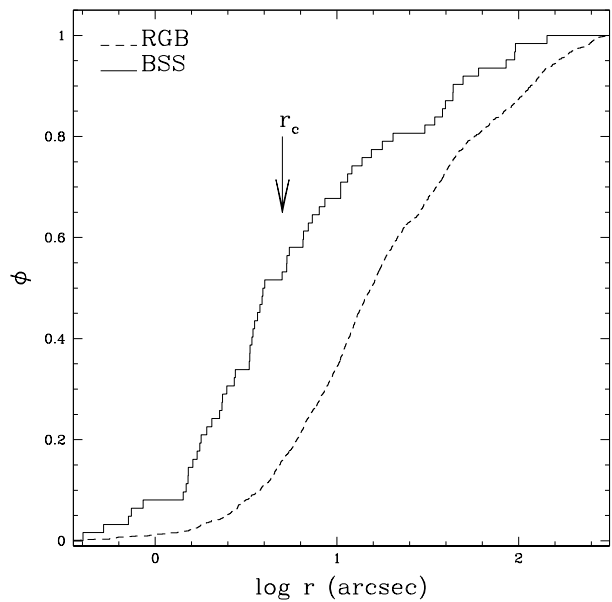}
\caption[]{Cumulative radial distribution of BSSs (solid line) and RGB stars
(dashed line) as a function of the projected distance from the cluster
$C_{\rm grav}$ for the combined HR+WF sample. The location of the cluster tidal
radius is marked by the vertical arrow.}
\label{acumul}
\vspace{0cm}
\end{figure}

\subsubsection{Field Decontamination}
The close inspection of the CMDs (in particular the diagram shown in
the right--hand panel of Fig.~\ref{cmdoptwfi}) suggests that the selected
populations can be affected by contamination from field
stars. As expected, the relative contribution of background star
increases progressively at large distance from the center,
where the cluster population significantly decreases and the Galaxy field
population is dominant. In order to quantitatively evaluate this effect
we took advantage of the large FOV of the WF sample.
Hence, we have used the outermost portion ($r\geq900^{\prime\prime}$) 
of the WFI data set, well beyond the tidal radius (see Fig.~\ref{fov}), 
to statistically subtract the background contribution.  
Accordingly, we have estimated the area of this 
region ($\sim 500$ arcmin$^{2}$)
and counted the number of stars lying within the corresponding BSS, HB and RGB 
selection boxes shown in Fig.~\ref{cmdoptwfi}. The estimated contamination 
is roughly 0.008, 0.3 and 0.53 stars/arcmin$^{2}$ for 
BSSs, HB and RGB stars, respectively. 

\subsection{The BSS Radial Distribution}
\label{radialdist}
The cumulative radial distribution for the 62 bright-BSSs and the 1675 
RGB stars are plotted
in Fig.~\ref{acumul} as a 
function of their projected distance from the cluster center. 
It is evident from the plot that the BSSs (solid line) are more centrally 
concentrated than RGB stars (dashed line). In fact, approximately $50\%$ of
the BSSs are located in the 
first $\sim 4^{\prime\prime}$ ($0.75r_{\rm c}$), while
only $\sim 10\%$ RGB stars are enclosed within the same distance. 
Hence, there is a preliminary evidence that the radial distribution of the 
BSS is quite different from that of the normal cluster populations.
We used a Kolmogorov-Smirnov (KS) test in order to check the 
null hypothesis that the radial distributions of the BSS and RGB populations
are identical, finding
a probability of $\sim 10^{-6}\%$. 
Thus, the BSS population in M75 has a different radial 
distribution than the RGB stars with more than $99\%$ probability.

\begin{deluxetable}{lllllc}
\footnote{Listed values are star counts before the field decontamination.
Numbers in parenthesis are the expected number of field stars.}
\tablecaption{Number Counts of BSSs, RGB, and HB stars, and Fraction of 
Sampled Luminosity in each annulus defined in section \ref{radialdist}.}
\tablewidth{0pt}
\tablehead{$r_{\rm i}^{\prime\prime}$&$r_{\rm e}^{\prime\prime}$&$N_{\rm BSS}$&$N_{\rm RGB}$&$N_{\rm HB}$&$L_{\rm samp}/L_{\rm samp}^{\rm tot}$\\}
\startdata
0   &  5  &33     &266     & 77     &0.16 \\
5   & 10  & 9     &311     & 82     &0.20 \\
10  & 15  & 5     &248     & 65     &0.15 \\
15  & 40  & 7     &400     & 132    &0.22 \\
40  & 60  & 3     &123(1)  & 44     &0.08 \\
60  & 90  & 2     &86(1)   & 25(1)  &0.06 \\
90  &150  & 3     &137(7)  & 50(4)  &0.09 \\
150 &300  & 0     &104(31) & 45(18) &0.04
\enddata
\label{tablecounts}
\end{deluxetable}

\begin{figure}
\includegraphics[width=8.5cm]{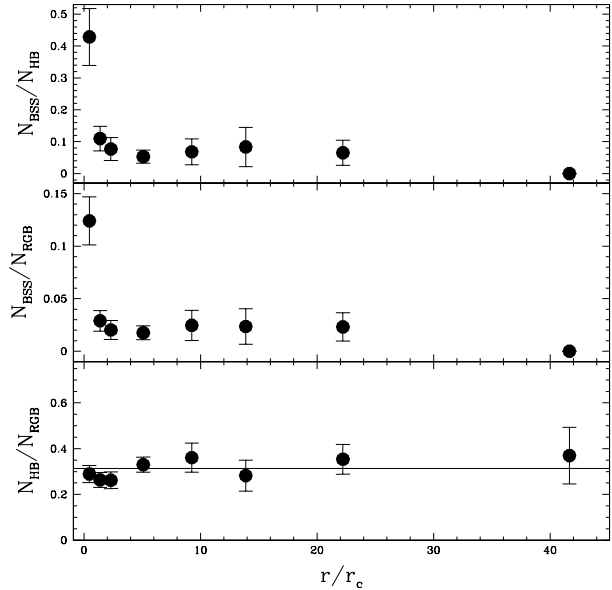}
\caption[]{Radial distribution of the population 
ratios $N_{\rm BSS}/N_{\rm HB}$, $N_{\rm BSS}/N_{\rm RGB}$ 
and $N_{\rm HB}/N_{\rm RGB}$ (top, middle and bottom, respectively)
as a function of the radial distance from the cluster center in core radius 
units.}
\label{spefre}
\vspace{0cm}
\end{figure}

For a more quantitative analysis of the BSS distribution, we
have followed the procedure described in previous papers
\cite[see e.g.,][]{ferraro04,sabbi04} and computed 
the specific frequencies $F^{\rm BSS}_{\rm RGB}=N_{\rm BSS}/N_{\rm RGB}$ and
$F^{\rm BSS}_{\rm HB}=N_{\rm BSS}/N_{\rm HB}$, where 
$N_{\rm BSS}$, $N_{\rm RGB}$ and$N_{\rm HB}$ represent the number of BSSs, 
RGB and HB stars, respectively. In doing this, the surveyed 
area has been divided into a set of concentric annuli and the number of BSSs,
RGB and HB stars contained in each annulus were counted. 
Star counts and the corresponding number of field stars (in parenthesis) for
each population in each annulus are listed in Table~\ref{tablecounts}.
The specific frequencies as a function of the projected
distance from $C_{\rm grav}$ are shown in Fig.~\ref{spefre}. 
For sake of completeness, the bottom panel plots the behavior of the 
ratio $N_{\rm HB}/N_{\rm RGB}$, showing that the selected reference 
populations share the same radial distribution. This farther demonstrates 
that the BSS radial distribution is peculiar independently of the adopted
reference population. Indeed, both $N_{\rm BSS}/N_{\rm HB}$ 
and $N_{\rm BSS}/N_{\rm RGB}$ show clearly a unimodal trend: the highest
value is reached in the innermost annulus 
($F^{\rm BSS}_{\rm RGB}\sim0.13$ and $F^{\rm BSS}_{\rm HB}\sim0.42$), 
and then the distribution decreases steeply, 
remaining flat as $r$ increases. 

\begin{figure}
\includegraphics[width=8.5cm]{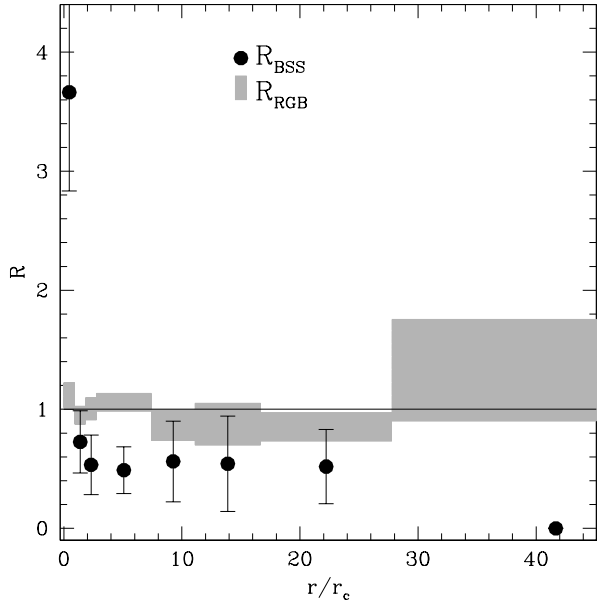}
\caption[]{Radial distribution of the BSS (large filled circles) and RGB 
(gray regions) relative frequencies as defined in eq. (1) and (2) as a 
function of the radial distance in units of core radius 
($r_{\rm c}=5^{\prime\prime}$). The vertical width of the gray regions 
correspond to the error bars.}
\label{relfre}
\vspace{0cm}
\end{figure}

As a final check, we have computed the double normalized  ratios (or relative
frequency) for BSSs and RGB stars, i.e, the ratio of the 
fractional number of stars in a given evolutionary phase (BSS, RGB)
to the fractional sampled luminosity in each annulus, as defined by 
\cite{ferraro93}:
\begin{equation}
R_{\rm BSS}=(N_{\rm BSS}/N^{\rm tot}_{\rm BSS})/(L_{\rm samp}/(L_{\rm samp}^{\rm tot}),
\end{equation}
\begin{equation}
R_{\rm RGB}=(N_{\rm RGB}/N^{\rm tot}_{\rm RGB})/(L_{\rm samp}/(L_{\rm samp}^{\rm tot}).
\end{equation}
The sampled luminosity in each annulus has been obtained by
integrating the best--fitting King profile 
(Fig.~\ref{densprof}), and appropriately scaling to the effective area
covered by the observations in each annulus.
The fraction of sampled to total luminosity is 
reported in the last column of Table~\ref{tablecounts}.
The relative frequency of the BSSs is plotted against radius in 
Fig.~\ref{relfre}, together with that of RGB stars for comparison. 
As the figure shows, $R_{\rm RGB}$ is essentially uniform over 
the surveyed area, with a value close to unity. 
This behavior is exactly what is expected on the basis of the 
stellar evolution theory, since the number of stars in any 
post-MS phase scales linearly with the total 
luminosity of the parent population \citep{renzini86}. This 
implies that a ``normal'' (i.e, non segregated) post-MS 
population scales as the cluster luminosity at any distance from the center, 
exactly as found for RGB stars.
In contrast, BSSs follow a different radial distribution.
Their relative frequency reaches a maximum value at the center of the
cluster and decreases to a minimum value at a distances $\sim5r_{\rm c}$
($\sim 25^{\prime\prime}$ from $C_{\rm grav}$), and remains
approximately constant outwards.
This behavior fully confirms that the BSS radial distribution in
M75 is not bimodal.

\section {Discussion and Conclusions}
We have studied the brightest portion ($m_{\rm F225W}\le21.3$, roughly 
corresponding to $m_{\rm F555W}\le20.2$) of the BSS population in M75. 
We have identified a total of 62 objects over a region that covers
the entire cluster extension. 
The BSS population has been found to be highly 
segregated toward the cluster center: indeed, approximately 60\%
of the entire BSS population is found within the cluster core, while
only $\sim 10-15\%$ of RGB and HB stars are counted in the same region. 
This suggests a significant overabundance of BSSs in the center, as also
confirmed by the fact that the BSS relative frequency ($R_{\rm BSS}$)
within $r_{\rm c}$ is roughly 3.5 times larger than expected for a normal
(non-segregated) population on the basis of the sampled light (see
Fig.~\ref{relfre}).  This value is one of the largest ever measured in the GCs 
studied so far with a similar approach. For example it is larger than 
that measured in M79 \citep[$\sim2.9$;][]{lanzoni07a}, 
M3 \citep[$\sim1.5$;][]{ferraro97}, 47~Tuc \citep[$\sim2.6$;][]{ferraro04}, 
and M5 \citep[$\sim2.7$;][]{lanzoni07b}.  

Notably, at odds with most of those  clusters, no
significant upturn of the distribution at large radii has been
detected in M75. Indeed the distribution found here is similar to 
that discovered by \cite{lanzoni07a} in M79: both these clusters 
show a central peak and a rapidly decreasing BSS specific 
frequency (within a few $r_c$), 
without any evidence of a rising branch in the outskirts.   
As discussed in \cite{lanzoni07a} for the case of M79, 
the absence of an external upturn in the BSS radial
distribution is not an effect of low statistics, since other clusters
show the external upturn in spite of a BSS population smaller or
similar to that measured in M75 \cite[see the case of NGC~6752 by][] 
{sabbi04}.
   
\begin{figure}[!t]
\includegraphics[width=8.5cm]{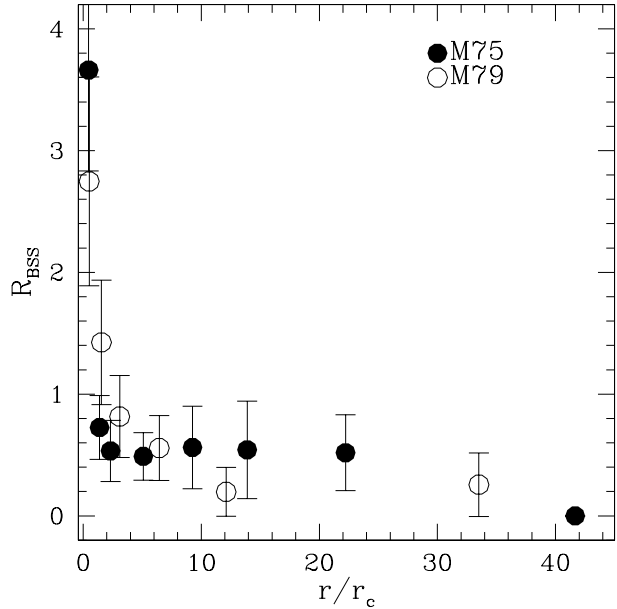}
\caption[]{Radial distribution of the double normalized ratio $R_{\rm BSS}$
for M75 (filled circles) compared to that measured in M79 (open circles)
by \cite{lanzoni07a}, plotted as a function of the radial distance from the
cluster center in core radius units.}
\label{relfrecomp}
\vspace{0cm}
\end{figure}

In order to  further explore the similarity between the 
BSS distribution in M75 and M79, we directly compare them in 
Fig.~\ref{relfrecomp}. 
As can be seen, although the behavior of the two distributions is
qualitatively similar, the decreasing trend of $R_{\rm BSS}$ in M75
appears to be 1) significantly sharper in the central region, where it
decreases by a factor of 5 (from 3.4 to 0.7) within just 1 $r_{\rm c}$
from the cluster center, and 2) somehow less steeper in the external 
regions, where it possibly define a sort of plateau. 

Dynamical simulations \cite[see][]{mapelli04,mapelli06,lanzoni07a}
have been used  to derive some hints about the origin of the BSS radial 
distribution. For example they demonstrated that the external rising 
branch of the BSS radial distribution observed in many GCs (M3, 47~Tuc, 
NGC~6752 and M5) cannot be due to BSSs which originate in the core and then 
kicked out in the outer regions. Instead it may be the consequence of binary
systems evolving in isolation in the cluster outskirts.  
In particular the measure of the so-called radius of avoidance 
($r_{\rm avoid}$, which indicates the distance at which the heavy 
stars like binaries have sunk to the cluster core) 
has been used to measure the efficiency of the dynamical friction in the
cluster. 
Following the procedure describe by \cite{beccari06}, we have derived
$log\rho_0=5.0\,M_\odot\,{\rm pc}^{-3})$ for the central density of
M75. Moreover, adopting the best--fit King model and assuming a 
velocity dispersion $\sigma_{0}=10.3$ km s$^{-1}$ \citep{mclaughlin05}, 12 
Gyr as the cluster age and the aforementioned computed 
value for the the central density, we have used the dynamical 
friction formula \citep[see e.g., eq.~1 in][]{mapelli06} to estimate 
$r_{\rm avoid}$ for M75, which is $\sim 17 r_{\rm c}$.
Hence, on the basis of this theoretical estimate, we would expect
a rising branch of the BSS population in the bins with $r
>r_{\rm avoid}$ $\sim 17 r_{\rm c}$, which is instead not observed. We note
that only 3 BSS have been observed in the annulus between 
$90^{\prime\prime}<r<150^{\prime\prime}$, and that 6 are needed to 
increase the $R$ value to 1 in Fig.~\ref{relfre}.
The situation is similar in the most external bin 
$150^{\prime\prime}<r<300^{\prime\prime}$, where 2 BSS are expected and 
0 are observed. Admittedly these are very small numbers largely
dominated by statistical fluctuations. However,
from a purely observational point of view, it is quite unlikely that we 
have lost a total of 5 BSS in such uncrowded region. On the other hand,
we the estimate of $r_{\rm avoid}$
is very sensitive to velocity dispersion ($\varpropto \sigma^{3}$),
which in the case of M75 has an uncertainty $\sim 20 \%$, meaning that
$r_{\rm avoid}\sim24r_{\rm c}$ may be achieved within the errors. 
It becomes clear that more accurate estimates of dynamical parameters 
are needed to better clarify this point.

The present study further demonstrates that BSS surveys covering the full
radial extent of GCs are powerful tools to determine how common
bimodality is and its connection to the cluster dynamical history.
A survey of physical and chemical properties for a  representative
number of BSSs in M75 as those performed by \cite{ferraro06b} in 47~Tuc 
and \cite{lovisi11} in M4 should clarify the 
formation processes of these stars.

\acknowledgments{This research is part of the project COSMIC-LAB
funded by the European Research Council (under contract ERC-2010-AdG-267675). 
The financial contribution of the Italian Istituto Nazionale di Astrofisica 
(INAF, under contract PRIN-INAF 2008) and the Agenzia Spaziale Italiana 
(under contract ASI/INAF/I/009/10) is also acknowledged. 
RTR is partially supported by STScI grant GO-11975-05.
We thank M. Catelan for providing us useful photometric data of M75 and
the anonymous referee for useful suggestions which improved the presentation
of the paper.}


\clearpage

\end{document}